\documentclass[11pt]{article}
\begin{document}
\begin{large}
\begin{titlepage}

\vspace{0.2cm}
\title{Top-charm associated production at hadron colliders
in the standard model with large extra dimensions
\footnote{Supported by National Natural Science Foundation of
China.}} \vspace{3mm}

\author{{ Zhou Hong$^{2}$, Ma Wen-Gan$^{1,2}$, and Zhang Ren-You$^{2}$ }\\
{\small $^{1}$ CCAST (World Laboratory), P.O.Box 8730, Beijing 100080, P.R.China} \\
{\small $^{2}$ Department of Modern Physics, University of Science and Technology}\\
{\small of China (USTC), Hefei, Anhui 230027, P.R.China} }

\date{}
\maketitle
\baselineskip=0.36in

\baselineskip=2in

\begin{abstract}
The precise calculations are carried out on the flavor changing
neutral current couplings in the process $pp \rightarrow gg
\rightarrow t\bar{c}(\bar{t}c)$ at the large hadron collider(LHC)
and very large hadron collider(VLHC) in both frameworks of the
minimal standard model(MSM) and its extension with extra
dimensions. We find that the effects from the large extra
dimensions can enhance the total cross section up to about several
hundred times as that in the MSM, quantitatively.
\end{abstract}

\vskip 4cm

\baselineskip=0.3in

{\large\bf PACS: 04.50.+h, 12.60.-i, 13.85.-t

  }

\vfill \eject
\end{titlepage}
\vskip 4cm

\par
It is known that the minimal standard model(MSM) has achieved a
great success under the electroweak energy scale, $m_{EW}$, which
is about $10^2~GeV$ energy scale \cite{RPSM}. But there still
exist several problems, such as hierarchy problem, which can not
been solved in this theory. There are many extensions of MSM being
proposed to solve these problems, such as supersymmetric models.
Recently, a new framework for solving the hierarchy problem, which
does not rely on either supersymmetry or technicolor, was
established by Arkani-Hamed, Dimopoulos and Dvali(ADD) \cite{ADD},
They predicted that if we confine the matter fields to our
4-dimensional world, the size of extra dimensions could be large
enough to be detectable. In this model, the gravitation and gauge
interactions are united at the weak scale, $M_D$, which is the
only fundamental scale in this theory, although it is now a
model-dependent parameter, and no new low-energy predictions can
be derived. This has the exciting implication that future
high-energy collider experiments can directly probe the physics of
quantum gravity. Above $TeV$ energy scale, completely new
phenomena could emerge as resonant production of the Regge
recurrences of string theory or excitations of Kaluza-Klein modes
of ordinary particles \cite{IKM}.
\par
There are stringent experimental constraints against the existence
of tree-level flavor changing scalar interactions (FCSI's)
involving the light quarks. This leads to the suppression of the
flavor changing neutral current(FCNC) couplings, an important
feature of MSM, which is explained in terms of the
Glashow-Illiopoulos-Maiani(GIM) \cite{GIM}. The advantage of
examining top quark physics than other quark physics is that one
can directly determine the properties of top quark itself and does
not need to worry about no-perturbative QCD effects which are
difficult to attack because there exist no top-flavored hadron
states at all. The properties of top quark makes the study on the
FCNC couplings to be one of important fields in top physics. The
FCNC couplings correlated with top quarks can be probed either in
rare decays of $t$-quark or via top-charm associated production
which appear at loop-level and offer a good place to test quantum
effects of the fundamental quantum field theory. In models beyond
MSM, new particles may appear in the loop and have significant
contributions to flavor changing transitions. Any positive
experimental observation of FCNC existence deviated from that in
the MSM would unambiguously pronounce the presence of new
physics\cite{All}. On the contrary, in case no deviation from the
SM is observed, they define in a quantitative way the strategy to
obtain lower bounds on the new physics energy scale.
\par
In this letter we shall investigate the process $pp \rightarrow
t\bar{c}(\bar{t}c)$ at hadron colliders in the MSM extension with
large extra dimensions. The cross section of the subprocess $pp
\rightarrow d\bar{d} \rightarrow t\bar{c}(\bar{t}c)$ is
proportional to $|V_{dt}^{*}\cdot V_{dc}|^2$, but that of the
subprocess $pp \rightarrow gg \rightarrow t\bar{c}(\bar{t}c)$ is
approximately proportional to $|V_{bt}^{*}\cdot V_{bc}|^2 $ for
$|V_{bt}^{*}\cdot V_{bc}|^2 \simeq |V_{st}^{*}\cdot V_{sc}|^2 $.
According to the latest data from Ref. \cite{particle},
$\frac{|V_{bt}^{*}\cdot V_{bc}|^2 }{|V_{dt}^{*}\cdot V_{dc}|^2}
\simeq 2.5\times 10^3$. The luminosity of gluon from proton at
hadron colliders is much larger than that of $d$-quark. Although
the process $gg \rightarrow t\bar{c}$ appears at one-loop level,
while the process $d\bar{d} \rightarrow t\bar{c}$ does at tree
level, we can still conclude that the cross section of the first
one will be much larger than that of the later process. We found
exactly that the cross section of $pp \rightarrow gg \rightarrow
t\bar{c}$ is almost 100 times large as that of $pp \rightarrow
d\bar{d} \rightarrow t\bar{c}$ after a precise numerical
computing. So we will focus only on the process $pp \rightarrow gg
\rightarrow t\bar{c}$ in the following investigation.
\par
There were some discussions about $t\bar{c}(\bar{t}c)$ production
at the LHC based on the THDM and the R-violating MSSM\cite{Zhou}.
The results of Ref. \cite{Zhou} show that the signals of the
$t\bar{c}(\bar{t}c)$ production induced by those models maybe
observable, but strongly depends on the chosen parameters. The
research on $t\bar{c}(\bar{t}c)$ production at $e^+e^-$ in the SM
was presented in Ref.\cite{CHChang}, in the THDM-III in
Ref.\cite{THDMIII} and in the MSSM in Ref.\cite{MSSMtc}. In this
work we consider the process $pp \rightarrow t\bar{c}(\bar{t}c)$
at hadron colliders in the MSM extension with large extra
dimensions. Within the framework of the MSM with extra dimensions,
the Feynman diagrams with graviton appear only in the $s$-channel
and the graviton presents as a propagator in the
$t\bar{c}(\bar{t}c)$ association production, but not appears in
loop. Then in this process the FCNC couplings originates still in
CKM mixing matrix rather than other new mechanics. Since the scale
of reaction energy at the LHC and VLHC has gone up to $14~TeV$ and
$100~TeV$, respectively, the effects of the new dimensions will be
expected emerge observably. There are eighteen diagrams including
the three diagrams involving graviton for the subprocess $gg
\rightarrow t\bar{c}(\bar{t} c)$ as shown in Fig.1.
\par
Here we present the relevant Feynman rules for our calculation
which can be extracted from Ref.\cite{ADD}. The relevant Feynman
rules are listed below,
$$
  for~~ Fig.2(a),~~~~~~~~~~~~~~~
  \frac{1}{2} \frac{i}{k^2-m^2}(\eta_{\mu \mu^{\prime}}
                  \eta_{\nu \nu^{\prime}}+
                  \eta_{\mu \nu^{\prime}}\eta_{\nu \mu^{\prime}}
                  -\frac{2}{D-2} \eta_{\mu \nu}\eta_{\mu^{\prime}
                  \nu^{\prime}}),
$$
$$
  for~~ Fig.2(b),~~~~~~~~~~~~~~~
i \frac{\kappa}{\sqrt V_{\delta}}[B_{\mu\nu\alpha\beta} m_A^2+
     (C_{\mu\nu\alpha\beta\rho\sigma}-
     C_{\mu\nu\alpha\sigma\beta\rho})k^{\rho}_1k^{\sigma}_2]\delta^{ab},
$$
$$
  for~~ Fig.2(c),~~~~~~~~
-i \frac{\kappa}{8 \sqrt V_{\delta}}[\gamma_{\mu}(p_1+p_2)_{\nu}+
                            \gamma_{\nu}(p_1+p_2)_{\mu}
                           -2 \eta_{\mu\nu}({\rlap/ p_1}+ {\rlap/ p_2}-2
                           m_{\psi})],
$$
respectively, where the flat matric
$\eta_{\mu\nu}=diag\{+1,-1,-1,-1\}$. $\delta=D-4$ is the number of
the extra dimensions. $V_{\delta}$ is the volume of the
compactified space. $B_{\mu\nu\alpha\beta}$ and
$C_{\mu\nu\alpha\beta\rho\sigma}$ are defined as below
\begin{eqnarray*}
B_{\mu\nu\alpha\beta} &=& \frac{1}{2} (\eta_{\mu\nu} \eta_{\alpha\beta} -
               \eta_{\mu\alpha} \eta_{\nu\beta} -
                \eta_{\mu\beta} \eta_{\nu\alpha}), \\
C_{\mu\nu\alpha\beta\rho\sigma} &=& \frac{1}{2} (
       \eta_{\mu\nu} \eta_{\alpha\beta} \eta_{\rho\sigma} - (
        \eta_{\mu\rho} \eta_{\nu\sigma} \eta_{\alpha\beta} +
      \eta_{\mu\sigma} \eta_{\nu\rho} \eta_{\alpha\beta} +
       \eta_{\mu\alpha} \eta_{\nu\beta} \eta_{\rho\sigma} +
         \eta_{\mu\beta} \eta_{\nu\alpha} \eta_{\rho\sigma} )).
\end{eqnarray*}
\par
For the s-channel virtual graviton exchange diagram, the sum over
all Kaluza-Klein modes has to be performed at the amplitude level
(The detailed deduction can be found in Refs.\cite{C9811} ). In
order to absorbing the factor $\frac{\kappa}{\sqrt V_{\delta}}$ of
the interaction vertex for which we have the equation
$\frac{\kappa}{\sqrt V_{\delta}}=\frac{2}{\bar{M}_P}$, where
$\bar{M}_P=M_P/\sqrt{8 \pi}=2.4 \times 10^{18} GeV$ is the reduced
Plank mass. The we can define
$$
D(s)=\frac{1}{\bar{M_P}^2} \sum_{n} \frac{1}{s-m_n^2}.
$$
Converting the sum into an integral which can be evaluated using dimensional
regularization, we have
$$
D(s)=-\frac{1}{M_D^{2+\delta}} \frac{S_{\delta-1}}{2} C \Lambda^{\delta-2}
$$
for $\delta>2$. Where $M_D \sim 10^2~GeV$ is the only fundamental
mass scale in the model, $S_{\delta-1}$ is the hypersurface of a
unit-radius sphere in $\delta$ dimensions, $\Lambda$ is the
ultraviolet cutoff, $C$ is an unknown coefficient. So $D(s)$ is
fixed by several uncertain parameter. Here we can postulate $C
\simeq 1$, $\Lambda \simeq M_D$ and have
$$
D(s)=-\frac{1}{M_D^4} \frac{S_{\delta-1}}{2}.
$$
In our calculation we take $M_D$ in the range of $100 \sim 500~GeV
$.
\par
In the numerical calculations the following input values of the
parameters have been employed \cite{particle}
\begin{eqnarray*}
&m_t=174.3~GeV,~~ m_c=1.25~GeV,~~m_b=4.2~GeV,~~m_s=0.12~GeV&\\
&m_W=80.42~GeV,~~\alpha=\frac{1}{128},~~\alpha_s=0.117.&
\end{eqnarray*}
We take the  "standard" parameterization of CKM matrix
\cite{particle} and chose
$$
s_{12}=0.221814,~~s_{23}=0.0459998,~~s_{13}=0.00500002,~~\delta_{13}=0.
$$
\par
The total cross section for parent process at $pp$ collider can be
obtained by folding the cross section of subprocess $\hat{\sigma}
(gg \rightarrow t\bar{c})$ with the gluon luminosity.

\begin{equation}
\sigma(s,pp \rightarrow gg \rightarrow  t\bar{c}+X)=
   \int_{(m_{t}+m_{c})^2/s} ^{1} d \tau \frac{d L_{gg}}{d \tau}
   \hat{\sigma} (gg\rightarrow t\bar{c} \hskip
3mm at \hskip 3mm \hat{s}=\tau s) \nonumber,
\end{equation}
where $\sqrt{s}$ and $\sqrt{\hat{s}}$ are the $pp$ and $gg$ c.m.s.
energies respectively, at the LHC $\sqrt s=14~TeV$ and the VLHC
$\sqrt s=100~TeV$, and $d L_{gg}/d \tau$ is the distribution
function of gluon luminosity, which is defined as
\begin{equation}
\frac{d L_{gg}}{d\tau}=\int_{\tau}^{1} \frac{dx_1}{x_1} \left[
f_g(x_1,Q^2)f_g(\frac{\tau}{x_1},Q^2) \right], \nonumber
\end{equation}
where $\tau = x_1~x_2$, the definitions of $x_1$ and $x_2$ are
from Ref.\cite{Zhou}, and in our calculation we adopt the CTEQ5L
parton distribution functions \cite{Martin}. The factorization
scale Q is chosen as the average of the final particles masses
$\frac{1}{2}(m_{t} +m_{c})$. The total cross section of $pp
\rightarrow t\bar{c}(\bar{t}c)+X$ is obtained by multiplying the
cross section of $pp \rightarrow t\bar{c}+X$ by factor 2. We have
the total cross section in the MSM $4.23 \times 10^{-3}fb$ at the
LHC and $0.105fb$ at the VLHC.
\par
The cross sections of parent process $pp \rightarrow gg
\rightarrow t\bar{c}+\bar{t}c+X$ at the LHC as the functions of
$M_D$ are depicted in Fig.3 with $\delta=3,5,7$, respectively. The
figure shows that when $M_D=100~GeV$, the cross sections are about
$1~fb$ and sharply decrease with the increament of  $M_D$ at
first, then approach asymptotically to the cross section in the
MSM after the point of $M_D=400~GeV$. In Fig.5 the curves describe
the relations between the cross sections of parent process $pp
\rightarrow gg \rightarrow t\bar{c}+\bar{t}c+X$ and $M_D$ at the
VLHC, which is almost the same feature as that at the LHC but the
cross sections at the VLHC are generally as several hundred times
as those at the LHC. So if we take $M_D=100~GeV$ and assume that
the LHC and the VLHC run with integrated luminosity of $L \simeq
300[fb]^{-1}$ in one year, there will be the order of $10^2 \sim
10^3$ events per year accumulated at the LHC and $10^4 \sim 10^5$
at the VLHC, respectively. In Fig.4 and Fig.6, we present the
cross section relative enhancement
$(\sigma-\sigma_{SM})/\sigma_{SM}$ by the virtual graviton
exchange as the functions of $M_D$ at the LHC and the VLHC,
respectively. Since $M_D$ appears in denominator of graviton
propagator, the cross section decreases with the increment of
$M_D$ as shown in the figures. We can figure out that the
correction from the large extra dimensions to the SM cross section
is approximately proportional to $M_D^{-4}$. The contribution of
$\delta$ origins mainly in the hypersurface of a unit-radius
sphere in $\delta$ dimensions. As we know,
$S_{\delta-1}=\pi^{\frac{\delta}{2}}/\Gamma(\frac{\delta}{2})$. We
have $S_{\delta-1}|_{\delta=3} <S_{\delta-1}|_{\delta=5}
<S_{\delta-1}|_{\delta=7}$, so the cross sections of $\delta=7$
are larger than those of $\delta=5(\delta=3)$ in the same
condition as shown in all figures.
\par
To summarize, we find that the the cross section of top-charm
associated production at hadron colliders is largely enhanced due
to large extra dimensions within the framework of the MSM with
large extra dimensions, and can be detectable both at the LHC and
the VLHC with the favorable parameters. This cross section
enhancement for the $t\bar{c}(\bar{t}c)$ association production
process is strongly related to the energy scale $M_D$. When $M_D$
is above $400~GeV$, the difference between the cross sections in
the MSM with and without large extra dimensions is indistinctly
demonstrated. The measurement of the cross section of $pp
\rightarrow gg \rightarrow t\bar{c}+\bar{t}c+X$ process can be
used to give the constraint on the fundamental energy scale $M_D$.

\par
\noindent{\large\bf Acknowledgements:}
\par
This work was supported in part by the National Natural Science
Foundation of China and a grant from University of Science and
Technology of China.


\vskip 10mm
\begin{flushleft} {\bf Figure Captions} \end{flushleft}

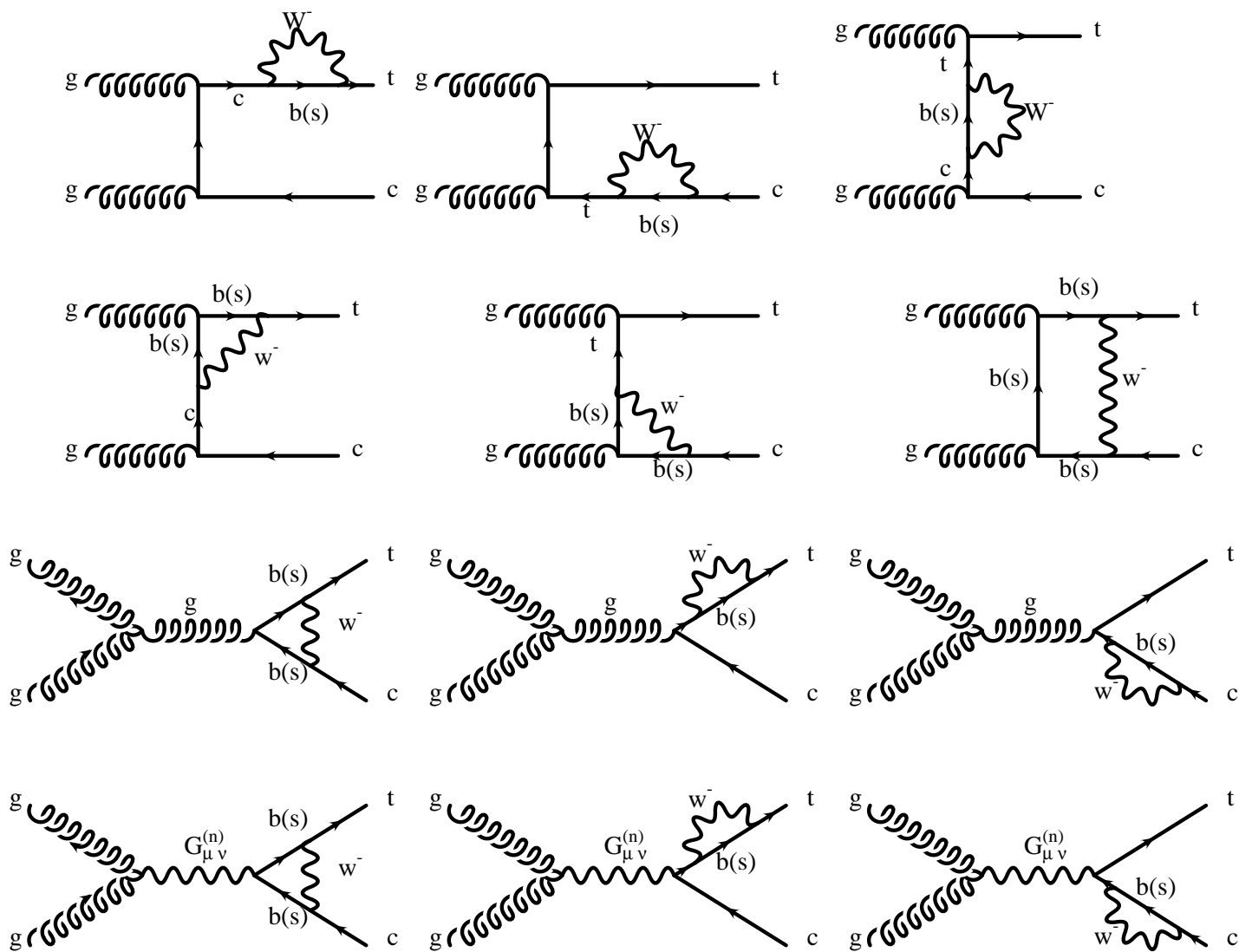
{\bf Fig.1} The Feynman diagrams of the subproecesses
$gg \rightarrow t\bar{c}+\bar{t}c$.

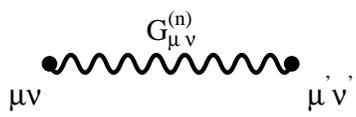
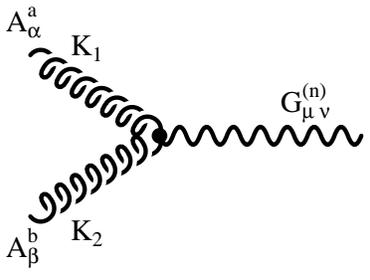
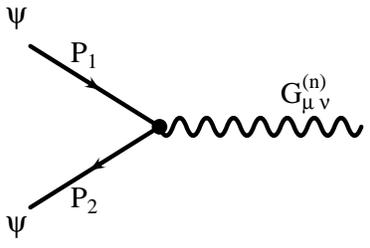
{\bf Fig.2} (a) The propagator of graviton. (b) The coupling
between the graviton and gluons. (c) The coupling between the
graviton and fermions.

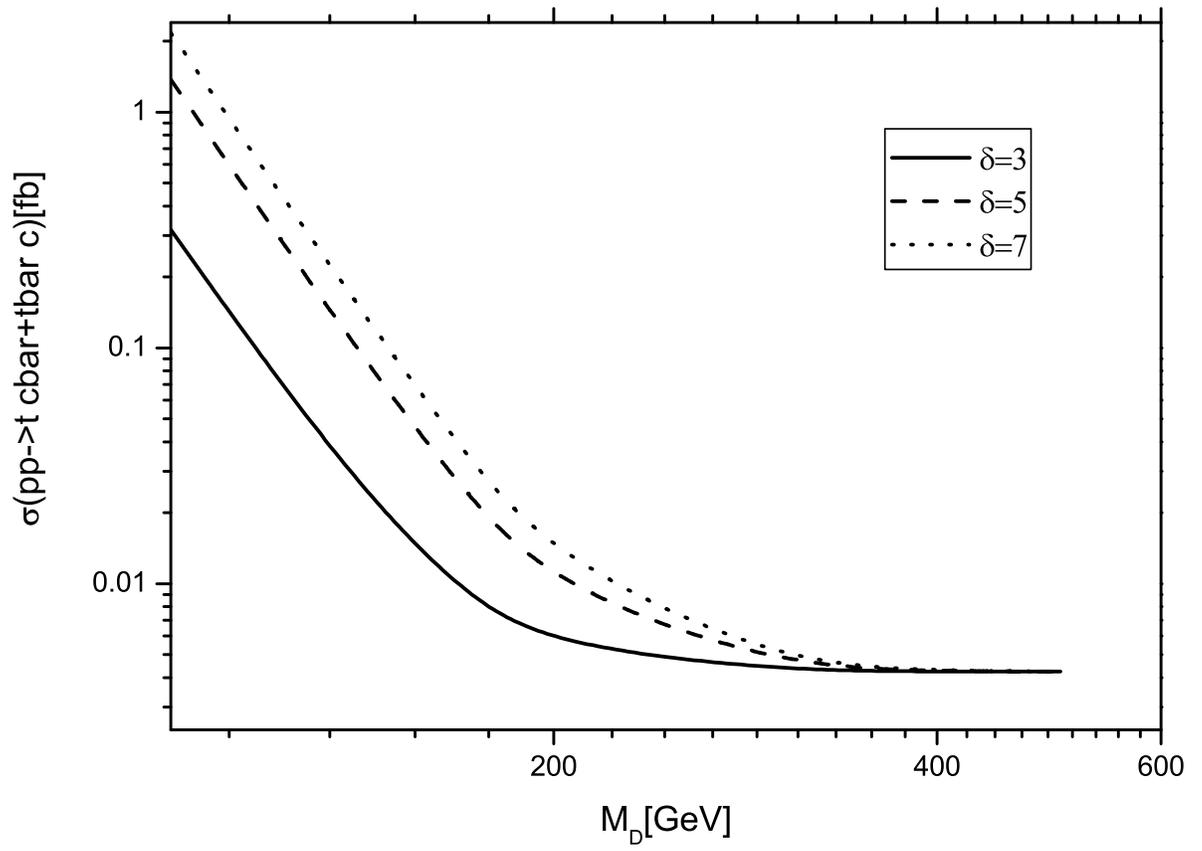
{\bf Fig.3} The cross sections of $pp \rightarrow gg \rightarrow
t\bar{c}+\bar{t}c+X$ at the LHC as the functions of the energy
scale $M_D$.

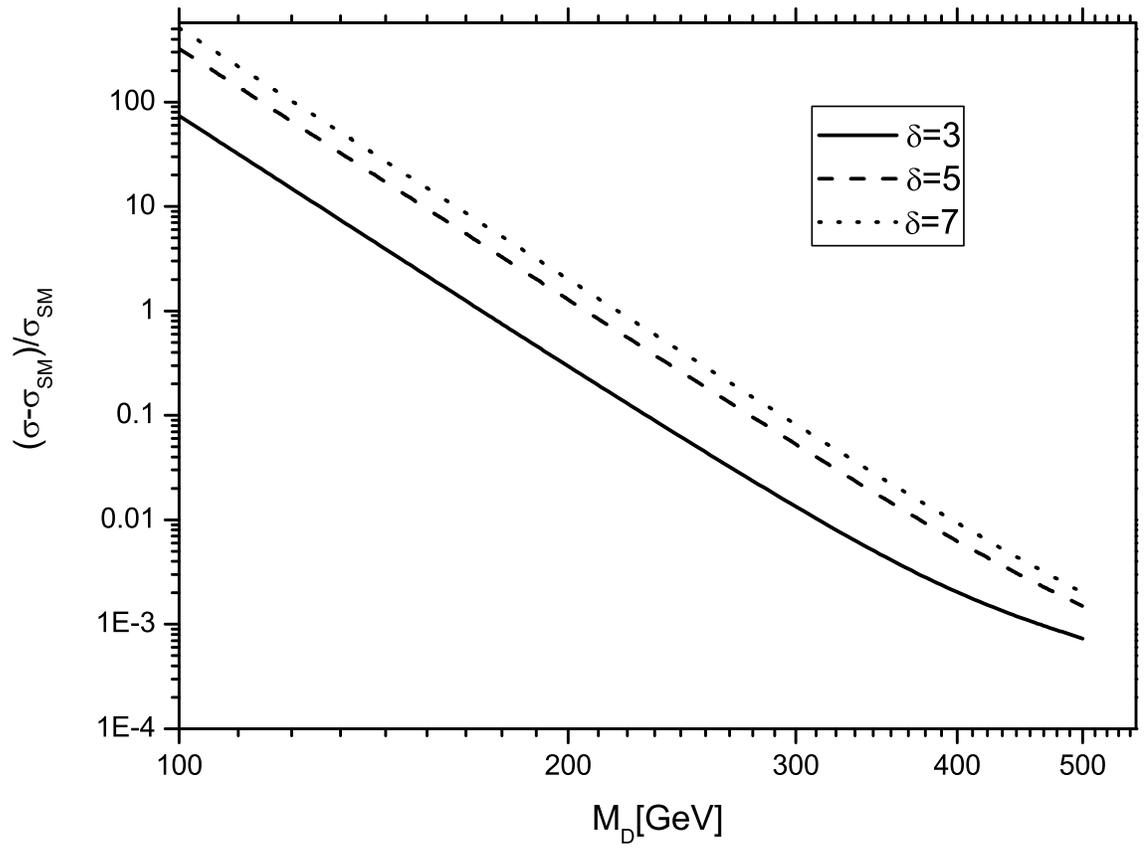
{\bf Fig.4} The enhancement to the cross sections of $pp
\rightarrow gg \rightarrow t\bar{c}+\bar{t}c+X$ at the VLHC as the
functions of the energy scale $M_D$.

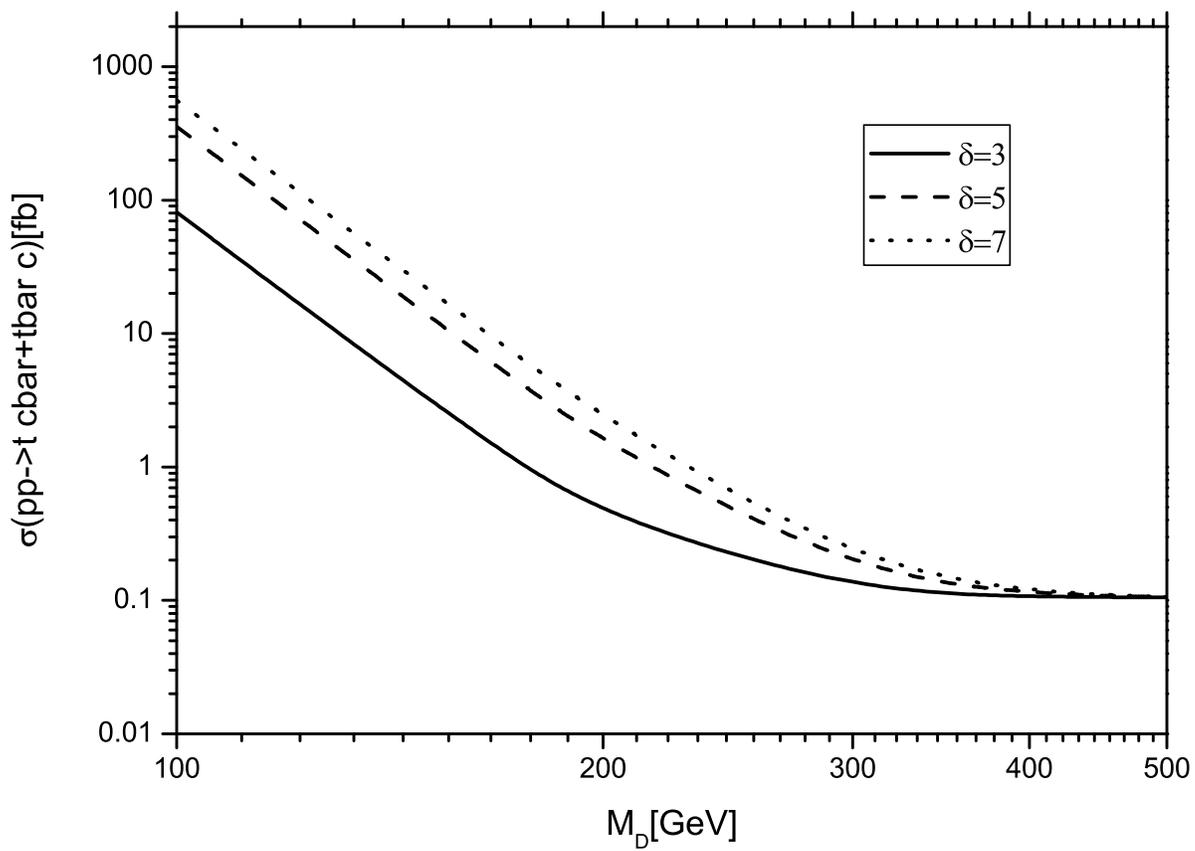
{\bf Fig.5} The cross sections of $pp \rightarrow gg \rightarrow
t\bar{c}+\bar{t}c+X$ at the LHC as the functions of the energy
scale $M_D$.

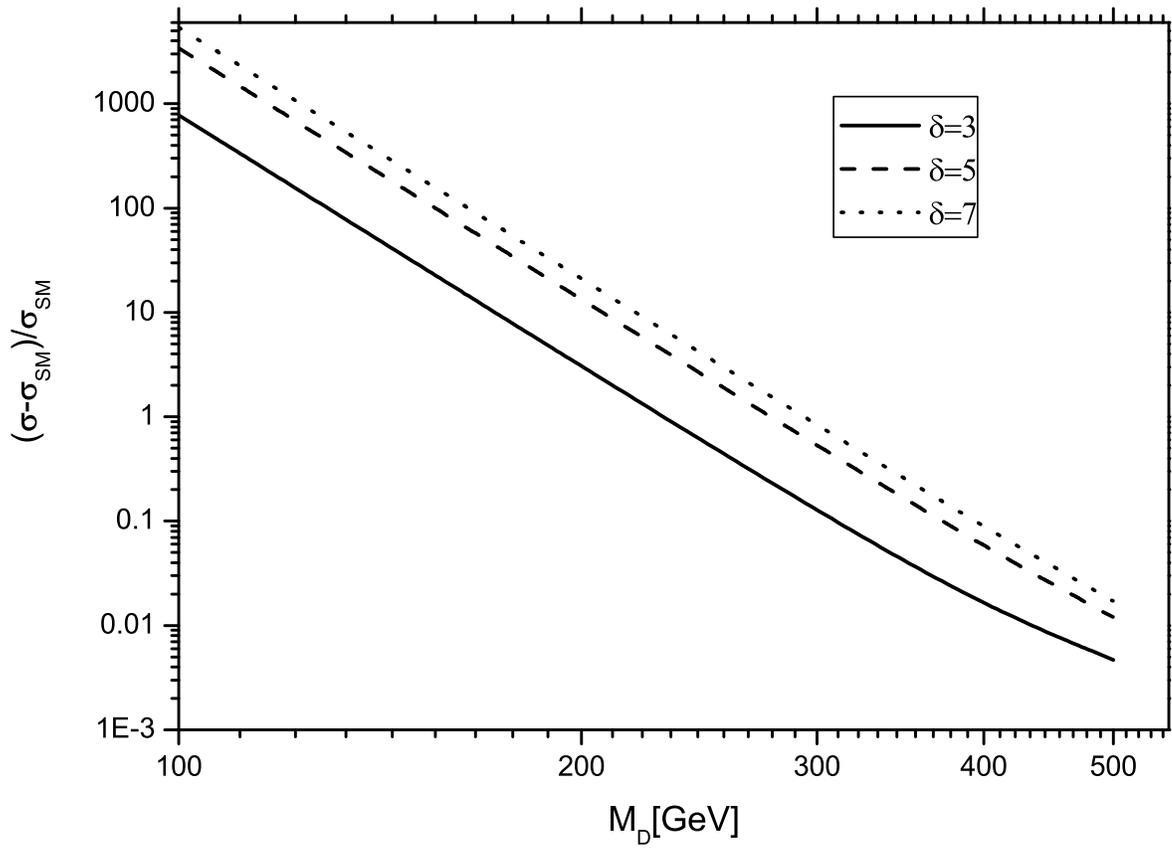
{\bf Fig.6} The enhancement to the cross sections of $pp
\rightarrow gg \rightarrow t\bar{c}+\bar{t}c+X$ at the VLHC as the
functions of the energy scale $M_D$.

\end{large}
\end{document}